\documentclass[twocolumn,article,notitlepage,floatfix,superscriptaddress,longbibliography,nofootinbib]{revtex4-1}

\usepackage{amsmath}
\usepackage{braket}
\usepackage{titlesec}
\usepackage{graphicx}
\usepackage{hyperref}
\usepackage[usestackEOL]{stackengine}
\usepackage{array}
\usepackage{tikz}
\usepackage[caption=false]{subfig}
\usepackage{bm}	
\usepackage{color}                     
\usepackage{xcolor}
\usepackage{epsfig}
\usepackage{amsmath} 
\usepackage{amssymb} 
\usepackage{longtable} 
\usepackage{float}
\usepackage{mathtools}
\usepackage{xparse}
\usepackage{hyperref}
\usepackage{times}
\usepackage[normalem]{ulem}
\usetikzlibrary{quantikz}
\graphicspath{ {./images/} }
\newcommand{\figref}[1]{Figure {\ref{#1}}}
\newcommand{\secref}[1]{Section {\ref{#1}}}

\newcommand{\tableref}[1]{Table {\ref{#1}}}
 
\renewcommand{\eqref}[1]{Eq. ({\ref{#1}})}

\begin{document}

\title{Introducing Non-Linear Activations into Quantum Generative Models}

\author{Kaitlin Gili}
\thanks{Both authors contributed equally to this work.}
\affiliation{University of Oxford, Oxford OX1 2JD }
\email{kaitlin.gili@physics.ox.ac.uk}

\author{Mykolas Sveistrys}
\thanks{Both authors contributed equally to this work.}
\affiliation{University of Oxford, Oxford OX1 2JD }

\author{Chris Ballance}
\affiliation{University of Oxford, Oxford OX1 2JD }
\affiliation{Oxford Ionics, Oxford, OX5 1PF}

\date{\today} 

\begin{abstract}

Due to the linearity of quantum mechanics, it remains a challenge to design quantum generative machine learning models that embed non-linear activations into the evolution of the statevector. However, some of the most successful classical generative models, such as those based on neural networks, involve highly non-linear dynamics for quality training. In this paper, we explore the effect of these dynamics in quantum generative modeling by introducing a model that adds non-linear activations via a neural network structure onto the standard Born Machine framework - the Quantum Neuron Born Machine (QNBM). To achieve this, we utilize a previously introduced Quantum Neuron subroutine, which is a repeat-until-success circuit with mid-circuit measurements and classical control. After introducing the QNBM, we investigate how its performance depends on network size, by training a 3-layer QNBM with 4 output neurons and various input and hidden layer sizes. We then compare our non-linear QNBM to the linear Quantum Circuit Born Machine (QCBM). We allocate similar time and memory resources to each model, such that the only major difference is the qubit overhead required by the QNBM. With gradient-based training, we show that while both models can easily learn a trivial uniform probability distribution, on a more challenging class of distributions, the QNBM achieves an almost 3x smaller error rate  than a QCBM with a similar number of tunable parameters. We therefore provide evidence that suggests that non-linearity is a useful resource in quantum generative models, and we put forth the QNBM as a new model with good generative performance and potential for quantum advantage.
\end{abstract}

\maketitle

\section{Introduction}\label{s:intro}

Recent advances in generative modeling have been integral to the success of many important data-driven tasks such as image generation \cite{huang2018introduction, karras2020analyzing}, drug discovery \cite{li2021quantum} and video prediction \cite{video_prediction}. Generative models can be defined as taking in data of some type, learning the defining features of the data, and outputting \textit{new} data of the same type and with the same features; in other words, learning a probability distribution from taking in a limited subset of samples from the underlying distribution. Learning complex probability distributions is a challenging problem for a classical computer, which is why improvements in quantum computing technology have prompted investigations into \textit{quantum} generative modeling \cite{2020Coyle, Liu_2018, Benedetti_2019, 2019zhu, 2021vqa, Rudolph2020GenerationOH, gen_mmd,kgili_gen, Zoufal_2019, qgen_review, hu_gan2019, Zoufal2019, huang2020QGANmnist}. This is partly because of how intuitively it can be embedded into a quantum computing framework, as probability distributions can be generated simply through measurements of quantum states. As hardware further improves, generative modeling has been identified as one of the leading candidates to achieve quantum advantage \cite{alcazar2021enhancing, han2018unsupervised, benedetti2019parameterized, gen_mmd, Benedetti_2019, Zoufal_2019, qgen_review}. 

While some quantum algorithms exhibit an exponential speedup over classical counterparts (e.g. Shor's, HHL, K-Means) \cite{Shor_1997, Harrow_2009, 2019sarma}, the potential quantum advantage in generative modeling is more subtle - it is claimed that quantum circuits are more expressive, so fewer resources are required (at least from the scaling point of view) to learn probability distributions of the same complexity \cite{2020Du}. One of the most promising quantum generative modeling families are Quantum Circuit Born Machines (QCBMs), where the quantum circuit is parameterized with tunable gates and the optimization process is conducted classically \cite{2020Coyle, Liu_2018, Benedetti_2019, alcazar2020classical, gen_mmd}. These models have demonstrated remarkable capabilities of modeling target probability distributions with high accuracy, and are one of the leading candidates to outperform classical neural networks as noise levels drop and qubit counts grow. 

One of the most fundamental differences between QCBMs and classical generative models is the presence (or absence) of linearity. Any classical neural-network-based generative model needs non-linear activation functions of neurons to learn non-trivial data \cite{Hopfield1999NeuralNA, Hinton2006ReducingTD}. In contrast, QCBMs generate probability distributions through projective measurements of quantum states, which have been evolved in an entirely linear fashion. Therefore, it is interesting to see whether introducing non-linearities to the quantum state evolution can improve the quantum generative model's ability to learn challenging data distributions.


\begin{figure*}
    \centering
    \includegraphics[width = \textwidth]{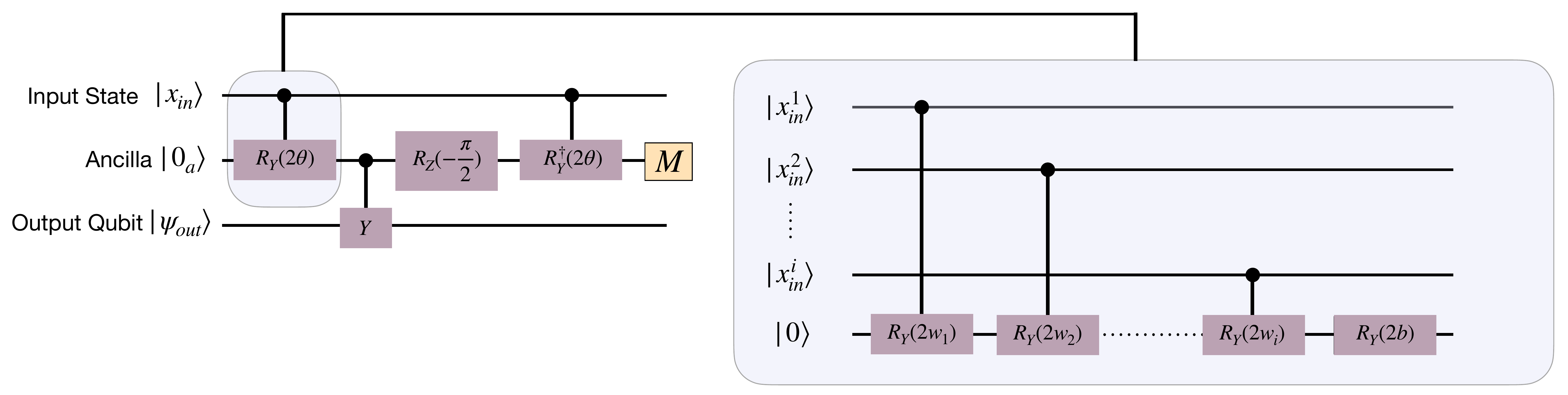}
    \caption{\textbf{An example of a single quantum neuron circuit} Left: Quantum Neuron circuit, connecting neurons (qubits) from previous layer to a single neuron in the next layer. If the ancilla measurement outcome is 0, the circuit has successfully applied $R_Y(2q(\theta))$ to the output qubit, and if the measurement outcome is 1, one needs to apply $R_Y(-\pi / 2)$ to the output qubit, a NOT gate to the ancilla, and apply the circuit again. Hence the Quantum Neuron circuit is a repeat-until-success circuit. Right: implementing $R_Y(2 \theta)$, where $\theta$ is a function of weights, biases, and input neuron values.}
    \label{fig:qneuron}
\end{figure*}

In this paper, we investigate the effect of adding non-linear activation functions to the Born Machine framework by introducing the Quantum Neuron Born Machine (QNBM) - a quantum generative model that mimics the connectivity and non-linear activation of classical neural networks. We expand upon previous work \cite{cao2017quantum} that utilizes repeat-until-success (RUS) circuits \cite{Paetznick_repeat,Bocharov_2015, rus_proof} as quantum analogues of neurons, and conduct an investigation into networks of these neurons as quantum generative models. We give a brief overview of how the QNBM scales in terms of tunable parameters, as well as the circuit depth, width, and qubit overhead. We then assess how various network structures (number of input and hidden neurons) affect the model's ability to express a non-trivial, cardinality-constrained 4-bit probability distribution \cite{garey79}. Finally, we compare the performance of our model to that of an all-to-all connected QCBM with $1$ parameterized single qubit gate layer and $1$ entangling layer, and demonstrate that our model outperforms the QCBM with gradient-based training when it comes to learning distributions with added complexity. Thus, we show the impact of introducing non-linear activations and neural network structure into the state evolution of quantum generative models.

\section{Quantum Generative Models}\label{s:model_overview}

In this section, we provide an overview of the QCBM algorithm and the Quantum Neuron (QN) building block, from which we construct the novel Quantum Neuron Born Machine (QNBM). Additionally, we provide insights into potential QNBM variations and the model's scaling limitations, prior to investigating its performance as a generative model.

\subsection{Quantum Circuit Born Machine (QCBM)}\label{sec2: qcbm}

The quantum component of a QCBM consists of a parameterized quantum circuit \cite{benedetti2019parameterized} that can generate a probability distribution by simply measuring its output state in the computational basis. Specifically, a discrete probability distribution with up to $2^{N_{out}}$ possible outcomes can be realized with an $N_{out}$ qubit circuit in the following way:

\begin{enumerate}
    \item Map every outcome to a bitstring $x$ with $N_{out}$ bits:  $x = \{0, 1\}^{N_{out}}$.
\item Map every bitstring $x$ to a quantum state $\ket{x} = \ket{x_1} \otimes \ket{x_2} ... \otimes \ket{x_i} \otimes ... \ket{x_{N_{out}}}, x_i \in \{ 0, 1\}$.
\item Run a quantum circuit with $N_{out}$ qubits for $N_{\text{shots}}$ and measure all qubits in the computational basis. The probability of outcome $x$ is the probability of the total quantum state to be found in $\ket{x}$ after measurement.
\end{enumerate}

In particular, if the quantum circuit is in the state $\ket{\psi}$ before measurement, then $P_{model}(x) = |\braket{x|\psi}|^2$ by the Born postulate - hence the name Born Machine \cite{2020Coyle, Liu_2018}. The QCBM learns the target probability distribution  $P_{target}(x)$ by varying the tunable circuit parameters $\theta$, until the final state $\ket{\psi} = U(\theta)\ket{0}$ generates a probability distribution $P_{model}(x)$ close enough to $P_{target}(x)$; here $U(\theta)$ is the unitary propagator of the quantum circuit.

Typically, one defines the similarity between the target and model distributions by the Kullback-Leibler (KL) divergence \cite{kl}:

\begin{equation}
KL = \sum_x P_{target}(x) \log(\frac{P_{target}(x)}{\max(P_{model}(x)}), 
\end{equation}

The KL is then optimized by a classical optimizer, either gradient-free or gradient-based. Ideal training ($P_{target} = P_{model}$) corresponds to $KL = 0$.

The most prominent quantum circuit structure (or \textit{ansatz}) is the hardware-efficient ansatz \cite{2019zhu} made of an entangling layer composed of two-qubit gates between each pair of qubits, and a parameterized layer composed of single-qubit Pauli X and Z rotations on each qubit. Both layers can be repeated several times with independent parameters, and the single qubit rotations are typically adapted for the specific depth and hardware. In keeping the ansatz structure fixed independent of depth, and taking each independent layer to include the single and two-qubit rotations, the total number of parameters becomes $(2N_{out} + {N_{out} \choose 2}) * L$ for a total number of output qubits $N_{out}$ and a total number of layers $L$.

\begin{figure*}
    \centering
    \includegraphics[width = \textwidth]{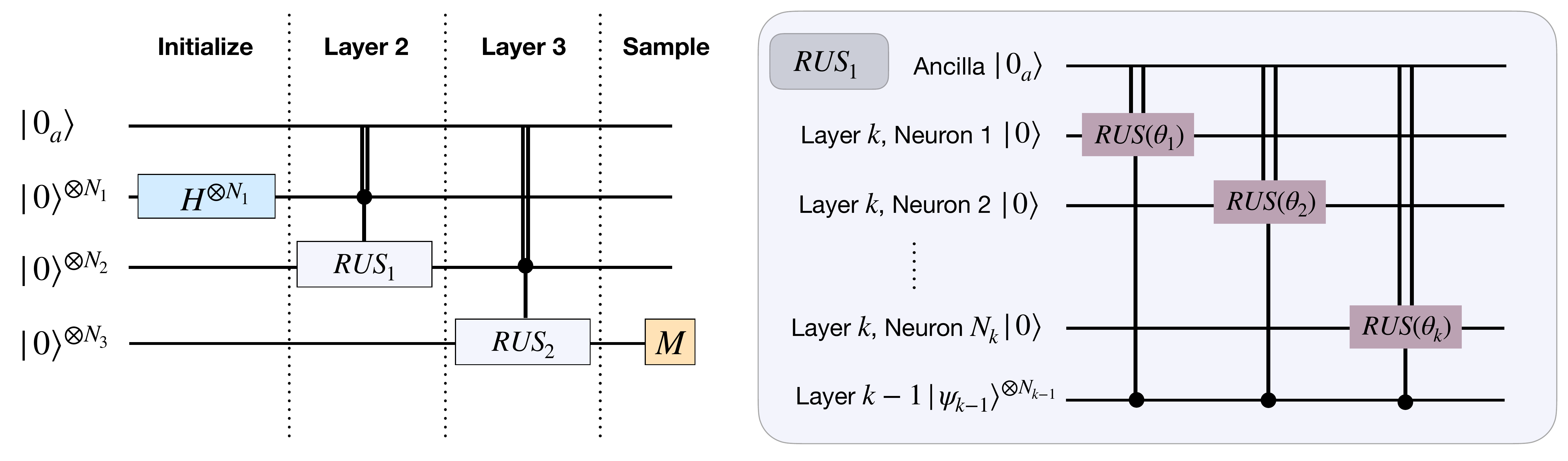}
    \caption{\textbf{Quantum circuit component of the Quantum Neuron Born Machine} Left: QNBM circuit with 3 layers, where RUS1 and RUS2 represent connections between layers. The input neurons are initialized in a uniform superposition, and the output probability distribution is generated by measuring the output layer. Right: connecting two layers of a QNBM. This amounts to running RUS quantum neuron circuits for every neuron in layer $k$.}
    \label{fig:qnbm}
\end{figure*}

\subsection{Quantum Neuron (QN)}\label{sec2:qneuron}

The basic building block of the QNBM is the quantum neuron circuit (shown in Figure \ref{fig:qneuron}),  which has previously been introduced as a potential building block for a quantum neural network \cite{cao2017quantum}. Each neuron in the network is assigned a qubit, so the QN circuit has an input register $\ket{x_{in}}$ representing the previous layer, an output qubit initiated in the state $\ket{\psi_{out}}$ representing the neuron of the next layer, and also an ancilla qubit, initially in $\ket{0_{a}}$. Suppose initially that $\ket{x_{in}}$ is a single bitstring, rather than a superposition of bitstrings. Then, just before the measurement (see \figref{fig:qneuron} left), the combined quantum state is: 
\begin{equation}
\begin{split}
    \ket{\psi} = \ket{x_{in}} \otimes ( \sqrt{p(\theta)} \ket{0_{a}} \otimes R_Y(2q(\theta)) \ket{\psi_{out}}  \\ + \sqrt{1-p(\theta)} \ket{1_{a}} \otimes R_Y(\pi / 2) \ket{\psi_{out}}),
\end{split}
\end{equation}
where $\theta$ is the weighted sum of neuron activations in the previous layer:
\begin{equation}
    \theta = w_1 x_1 + w_2 x_2 + ... + w_n x_n + b,
\end{equation}
${w_n} \in (-1;1)$ are the weights and $b \in (-1;1)$ is the bias. Notice the argument of the y-rotation is not $2 \theta$, but $2q(\theta)$: this is the activation function
\begin{equation}
q(\theta) = \arctan(\tan^2(\theta)).
\end{equation}
The activation function has a sigmoid shape, and is determined by the sequence of quantum gates applied. While we only use this activation function, we leave for future work the investigation of  gate sequences leading to other types of activation functions. In any case, by simply measuring the ancilla qubit to be $\ket{0_{a}}$ with a probability $p(\theta) > \frac{1}{2}$ \cite{cao2017quantum}, we end up in the right state for the next layer:
\begin{equation}
\ket{\psi} = \ket{x_{in}} \otimes \ket{0_a} \otimes R_Y(2q(\theta) \ket{\psi_{out}}.
\end{equation}
Importantly, when probability $1-p(\theta) >= \frac{1}{2}$ we will end up in the state: 
\begin{equation}
\ket{\psi} = \ket{x_{in}} \otimes \ket{1_a} \otimes R_Y(\pi/2) \ket{\psi_{out}},
\end{equation}
which can be returned to the pre-circuit state with a NOT gate on the ancilla and $R_Y(-\pi/2)$ applied to the output qubit. The process will then be repeated until the ancilla measurement yields $\ket{0_a}$, thus the Quantum Neuron circuit belongs to the class of Repeat Until Success (RUS) circuits \cite{Paetznick_repeat, Bocharov_2015}.

Importantly, if we allow the input register to be in a general state - as a \textit{superposition} of bitstrings $\sum_i \ket{x^i}$, then the circuit will successfully apply the appropriate transformations to each of the bitstrings in the superposition, resulting in the final state: 
\begin{equation}
\sum_i F_{i}\ket{x^{i}_{in}} \otimes \ket{0_a} \otimes R_Y(2q(\theta^{i})) \ket{\psi_{out}}. 
\end{equation}

Here, $F_{i}$ refers to an amplitude deformation in the input state during the RUS mapping. While these QN building blocks have been previously used for designing small-scale quantum networks for supervised learning tasks \cite{cao2017quantum}, here we modify the QN circuit for generative modeling. Specifically, we add parameterized non-linear transformations into the Born Machine framework.

\subsection{From QN to QNBM}\label{sec3:qnn}

So far we have only shown how a \textit{single} neuron in a layer can be connected to all neurons in the previous layer. Furthermore, connecting layer $k$ to layer $k-1$ in a feed-forward network is a trivial extension - one can repeat the QN circuit for every neuron in layer $k$ (as shown in \figref{fig:qnbm}). Note that we restrict this work to $k = 3$, where the number of neurons in the input, hidden, and output layers are denoted as $(N_{in}, N_{hid}, N_{out})$, respectively. Also, notice that the ancilla qubit can be reused for every RUS circuit, since after every circuit the ancilla is always in $\ket{0_a}$, and it is disentangled from all neurons. Thus, subsequent ancilla measurements do not affect already connected neurons. Likewise, connecting the \textit{whole} network simply amounts to repeating the above for every pair of neighbouring layers.

The quantum neural network proposed here is application-agnostic, and could, for example, be used in classification by adding data embedding and a cost calculation. To use it as a generative model, we make two alterations. First, the quantum state of the input layer: we initialize the first layer in a uniform superposition of bitstrings with Hadamard gates, as it is able to input some inherent structure into our circuit as an analogue to a classical prior distribution. Secondly, we only measure the qubits representing the output layer, in analogy to classical neural networks. The resulting probability distribution is then the output of the QNBM, and is optimized exactly the same way as in a QCBM, with the weights and biases acting as the free parameters.

\subsection{Variations \& Scaling}\label{sec3:scaling}

It's interesting to consider the effect of non-linear activation not just in a binary way (is non-linearity useful?), but also in a quantitative way (is \textit{more} non-linearity better?). It is, in fact, possible to alter the activation function and make it arbitrarily close to a step-like function, that acts as identity for $\theta < \pi / 4$ and as a NOT gate for $\theta > \pi / 4$ \cite{cao2017quantum}. This can be done by recursively implementing the RUS circuit such that one produces a nested activation function
\begin{equation}
q^{\circ n}(\theta) = q(q(...(q(\theta)).
\end{equation} 
However, this comes at a cost of having $n$ ancillas instead of one, as well $O(2^n)$ measurements and an increase in gates by a factor of $O(2^n)$. It is possible that the advantages of a more non-linear ($n > 1$) activation could outweigh the exponential resource requirements for larger datasets; however, in this paper we limit ourselves to a binary comparison of linear and non-linear models, and only use $n = 1$. 

As a quantum circuit, the QNBM has $O(E)$ two-qubit parameterized gates, $O(N)$ one-qubit parameterized gates, and $O(N)$ two-qubit and one-qubit fixed gates, where $E$ and $N$ are the number of total edges and nodes (neurons) in the network, respectively. The circuit gets longer if any Quantum Neuron subroutines need to be repeated; \cite{cao2017quantum} show, however, that the expected number of repetitions for each Quantum Neuron is bounded from above by $7$, regardless of total network size. 

Due to difficulties in simulating quantum circuits with classical control, we run each Quantum Neuron subroutine once (instead of repeating until success) and post-select experimental runs where every mid-circuit measurement produces the favorable outcome. To fairly quantify the model's time resources, we use the number of shots \textit{before} post-selection as the appropriate measure in our circuit simulations and model comparison.

Lastly, we note that the circuit also has $N + 1$ (or in the case $n > 1$, $N + n$) qubits, of which only $N_{out}$ neurons are measured to obtain the output. This results in a qubit overhead not present in other Born Machines. The larger total Hilbert space makes classical simulations of the QNBM longer than for a similarly-sized QCBM, but we do not expect the qubit overhead to increase the time resources of the model when realized experimentally. Nevertheless, we envision that for larger problem sizes, repeating-until-success is the only scalable way to use the QNBM.

\section{Results}\label{s:results}

In this section, we present our initial results from investigating the non-linear QNBM as a quantum generative model. First, we demonstrate the performance of various QNBM architectures, each containing different neuron sizes in the input $N_{in}$, hidden $N_{hid}$, and output $N_{out}$ layers. We use our numerical simulations to highlight some of the network patterns that lead to learning the data well. Secondly, we demonstrate the power of introducing non-linearity into state evolution of quantum generative models by comparing the learning abilities of a non-linear QNBM and the linear QCBM for both trivial and more challenging target probability distributions \cite{garey79}. 

\subsection{Investigating Network Design}\label{sec3:network-design}

\begin{figure}[!t]
    \includegraphics[scale=1.0, width=\linewidth]{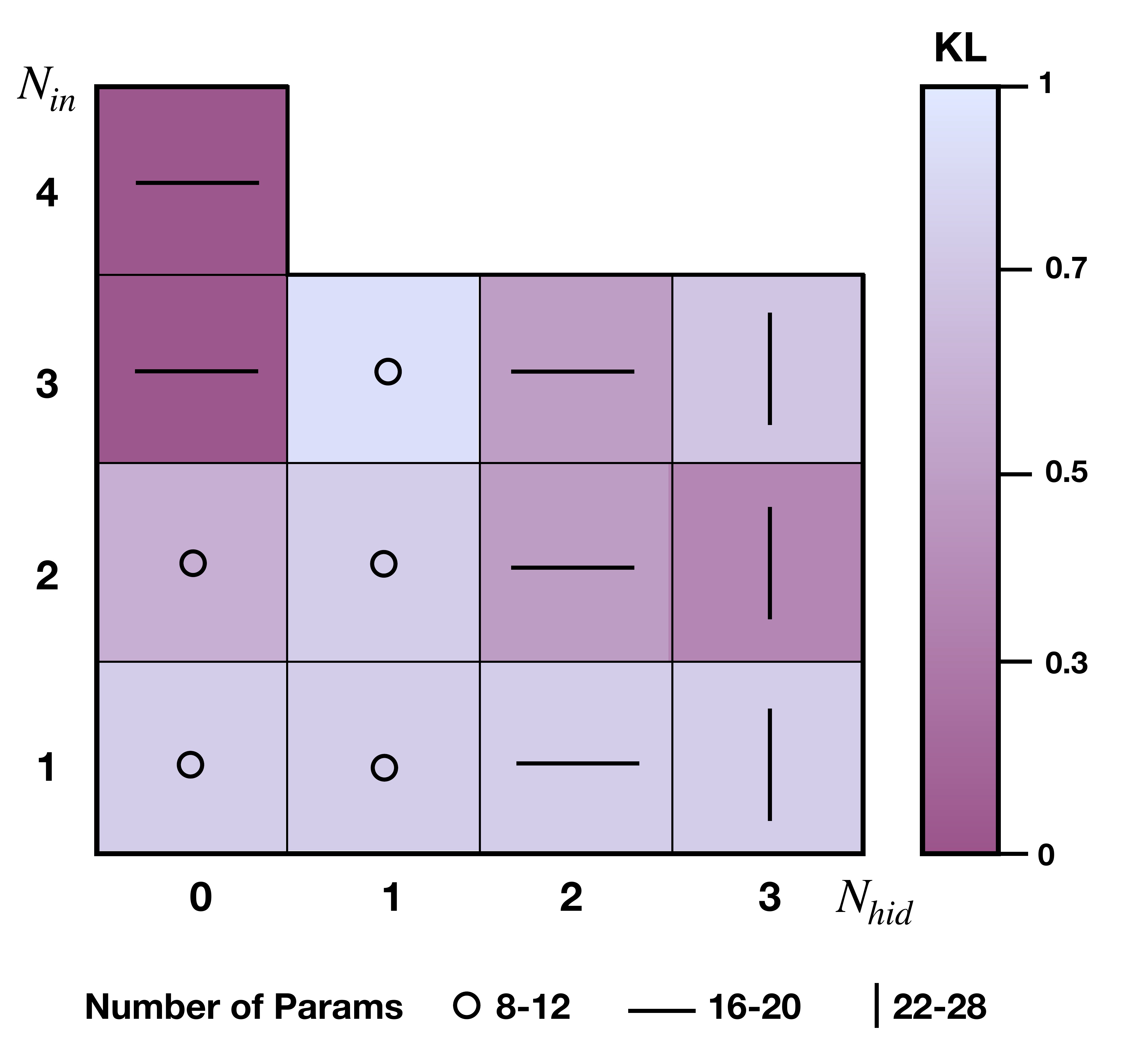}
\caption{\textbf{2D grid visualization of the KL Divergence vs. network layer size for a QNBM with a 4-dimensional output layer.} Each grid square corresponds to a specific number of input $N_{in}$ and hidden layers $N_{hid}$ in the network, where the color indicates the average KL cost value across 5 training seeds, such that the networks with darker squares reproduce the target distribution with a higher accuracy. We additionally highlight the number of tunable parameters associated with each network structure with a specific shape. The worst performance occurs when $N_{in} = 1$ (lightest shades), indicating that that the size of the input network is too small. However, we see that when increasing $N_{in}$, it is not necessarily advantageous to increase $N_{hid}$ or simply the number of tunable parameters. Overall, we achieve the best performance when $N_{hid} = 0$ and for $N_{in} = N_{out}$ or $N_{in} = N_{out} - 1$.}
\label{fig:network_trends}
\end{figure}

To obtain a first understanding of our QNBM as a generative model, we assess the model's ability to express a target distribution that is defined by a cardinality constraint over discrete bitstrings. Given a search space of $2^{N_{out}}$ discrete bitstring samples, we define a sub-space $\mathcal{S}$ of valid samples $x$ that have the desired cardinality $c = \lfloor \frac{N_{out}}{2} \rfloor$ (e.g. '0011', c = 2). The target distribution is uniform over these valid samples, and the goal of each QNBM is to generate high quality samples as a result of approximating the following distribution: 
\begin{equation}\label{data_prob}
P_{\mathcal{S}}(x) = \frac {1} {|\mathcal{S}|}, \forall x \in \mathcal{S}. 
\end{equation}

We note that while we use these seemingly toy distributions to benchmark our small-scale QNBMs, learning larger cardinality-constrained distributions is shown to be a difficult and important task in combinatorial optimization, for instance problems with financial datasets \cite{alcazar2020classical, gili2022evaluating}. Each model is trained with a KL Divergence cost function, comparing the overlap of the target distribution $P(x)_{target}$ and the generated distribution by the model $P(x)_{model}$ during each training iteration. Post-training, we assess the quality of the generated samples with the commonly used precision $P$ metric. We take $P$ to be the ratio of the number of valid samples $Q_{v}$ to the total number of queries generated from the model $Q$. To demonstrate perfect learning of the target distribution, we should see the following behavior: 
\begin{equation}\label{data_prob}
P = \frac {Q_{v}}{Q} \rightarrow 1.
\end{equation}
We show results for architectures with $N_{out} \in \{2, 3, 4\}$, spanning over the number of neurons in the input and hidden layer. Note that we restrict the scope of these results to investigating one hidden layer, and leave a study of adding additional layers to future work. All simulations are run with an Adam optimizer with a learning rate $\alpha = 0.2$ and a step size $\epsilon = 0.1$ for a specified number of maximum iterations $i_{\text{Max}}$. For $N_{out} = 2$, we set $i_{\text{Max}} = 200$ and for all other simulations $N_{out} > 2$, we set $i_{\text{Max}} = 500$ as we need to increase the training time in order to accommodate for introducing more tunable parameters into the model. We initiate the QNBMs with uniformly random weights and biases from $(-1,1)$. In order to demonstrate training robustness, we optimize each model over five different training seeds, and provide the average $P$ values along with their standard deviations for $Q = 10^4$ generated model samples. These results are shown in \tableref{table:network_span}. All simulations are run on IBMQ's Qiskit Aer Simulator with $N = 10^4$ shots.  

We display the performance of all the networks with $N_{out} = 4$ in Figure \ref{fig:network_trends}, where the networks are sorted by input and hidden layer size, as well as grouped into thirds by number of parameters: 8-12, 16-20, and 22-28. The data does not suggest significant trends from parameter number alone, but we do observe some trends when both parameter number and layer sizes are taken into account. Notably, the best performing networks have no hidden layer at all; the 3-0-4 and 4-0-4 achieved by far the best training, with over 95\% precision and KL $< 0.04$. Both of these networks were in the top two thirds in terms of parameter number - this suggests that while the extra parameters from hidden layers appear to be redundant for QNBMs, one needs an input layer similarly sized to the output layer to sufficiently parameterize the model; this point is further strengthened by the $N_{out} < 4$ data left out from the plot, but shown in Table \ref{table:network_span}. We believe the claims about optimal network arrangement have sufficient evidence for small-scale models, but we acknowledge the possibility that hidden layers could become useful for large-scale problems.

\begin{table}[htp]
\centering
\renewcommand{\arraystretch}{1.3}
{\scriptsize %
\begin{tabular}{||>{\centering}p{3cm} | >{\centering}p{1.5cm} | >{\centering}p{1cm} |
>{\centering}p{1cm} |
>{\centering}p{1cm} |
>{\centering}p{1cm} ||}
\hline
\multicolumn{1}{||c|}{$(N_{in}, N_{hid}, N_{out})$} & \multicolumn{1}{|c|}{
$P_{\text{num}}$} & \multicolumn{1}{|c|}{
$N_{\text{qubits}}$}
 & \multicolumn{1}{|c|}{\textbf{KL}} & \multicolumn{1}{|c||}{\textbf{Precision}} \\
\hline
\multicolumn{1}{||c|}{(1, 0, 2)} & \multicolumn{1}{|l|}{4} & \multicolumn{1}{|l|}{4} & \multicolumn{1}{|c|}{$0.0041 \pm 0.0001$} & \multicolumn{1}{|c||}{$0.9961 \pm 0.0002$} 
\\
\hline 
\multicolumn{1}{||c|}{(1, 1, 2)} & \multicolumn{1}{|l|}{6} &  \multicolumn{1}{|l|}{5} & \multicolumn{1}{|c|}{$0.06 \pm 0.02$} & \multicolumn{1}{|c||}{$0.94 \pm 0.02$} 
\\
\hline 
\multicolumn{1}{||c|}{(2, 0, 2)} & \multicolumn{1}{|l|}{6} & \multicolumn{1}{|l|}{5} &  \multicolumn{1}{|c|}{$0.102 \pm 0.002$} & \multicolumn{1}{|c||}{$0.903 \pm 0.002$} 
\\
\hline
\multicolumn{1}{||c|}{(1, 0, 3)} & \multicolumn{1}{|l|}{6} & \multicolumn{1}{|l|}{5} &  \multicolumn{1}{|c|}{$0.62 \pm 0.01
$} & \multicolumn{1}{|c||}{$0.59 \pm 0.03$} 
\\
\hline
\multicolumn{1}{||c|}{(1, 1, 3)} & \multicolumn{1}{|l|}{8} & \multicolumn{1}{|l|}{6} &  \multicolumn{1}{|c|}{$0.49 \pm 0.03$} & \multicolumn{1}{|c||}{$0.66 \pm 0.02$} 
\\
\hline 
\multicolumn{1}{||c|}{(1, 2, 3)} & \multicolumn{1}{|l|}{13}& \multicolumn{1}{|l|}{7} &  \multicolumn{1}{|c|}{$0.019 \pm 0.004$} & \multicolumn{1}{|c||}{$0.984 \pm 0.005$} 
\\
\hline 
\multicolumn{1}{||c|}{(2, 0, 3)} & \multicolumn{1}{|l|}{9} & \multicolumn{1}{|l|}{6} &  \multicolumn{1}{|c|}{$0.051 \pm 0.002
$} & \multicolumn{1}{|c||}{$0.951 \pm 0.002$} 
\\
\hline 
\multicolumn{1}{||c|}{(2, 1, 3)} & \multicolumn{1}{|l|}{9} & \multicolumn{1}{|l|}{7} &  \multicolumn{1}{|c|}{$0.446 \pm 0.003
$} & \multicolumn{1}{|c||}{$0.665 \pm 0.007
$} 
\\
\hline 
\multicolumn{1}{||c|}{(2, 2, 3)} & \multicolumn{1}{|l|}{15} & \multicolumn{1}{|l|}{8} &  \multicolumn{1}{|c|}{$1.0 \pm 1.0$
} & \multicolumn{1}{|c||}{$0.6 \pm 0.3
$} 
\\
\hline 
\multicolumn{1}{||c|}{(3, 0, 3)} & \multicolumn{1}{|l|}{12}  &\multicolumn{1}{|l|}{7} &  \multicolumn{1}{|c|}{$0.0517 \pm 0.0008
$} & \multicolumn{1}{|c||}{$0.946 \pm 
0.003$
} 
\\
\hline 
\multicolumn{1}{||c|}{(1, 0, 4)} & \multicolumn{1}{|l|}{8}  &\multicolumn{1}{|l|}{6} &  \multicolumn{1}{|c|}{$0.808 \pm 0.009
$} & \multicolumn{1}{|c||}{$0.460 \pm 0.007
$}
\\
\hline 
\multicolumn{1}{||c|}{(1, 1, 4)} & \multicolumn{1}{|l|}{10}  &\multicolumn{1}{|l|}{7} &  \multicolumn{1}{|c|}{$0.87 \pm 0.05
$} & \multicolumn{1}{|c||}{$0.44 \pm 
0.02$
} 
\\
\hline 
\multicolumn{1}{||c|}{(1, 2, 4)} & \multicolumn{1}{|l|}{16}  & \multicolumn{1}{|l|}{8} &  \multicolumn{1}{|c|}{$0.8 \pm 0.1
$} & \multicolumn{1}{|c||}{$0.49 \pm 
0.06$} 
\\
\hline 
\multicolumn{1}{||c|}{(1, 3, 4)} & \multicolumn{1}{|l|}{22} & \multicolumn{1}{|l|}{9} &  \multicolumn{1}{|c|}{$0.8 \pm 0.2
$} & \multicolumn{1}{|c||}{$0.53 \pm
0.06$} 
\\
\hline 
\multicolumn{1}{||c|}{(2, 0, 4)} & \multicolumn{1}{|l|}{12} & \multicolumn{1}{|l|}{7} &  \multicolumn{1}{|c|}{$0.59 \pm 0.03
$} & \multicolumn{1}{|c||}{$0.64 \pm
0.02$} 
\\
\hline
\multicolumn{1}{||c|}{(2, 1, 4)} & \multicolumn{1}{|l|}{11}  & \multicolumn{1}{|l|}{8} &  \multicolumn{1}{|c|}{$0.78 \pm 0.04$} & \multicolumn{1}{|c||}{$0.45 \pm
0.03$} 
\\
\hline
\multicolumn{1}{||c|}{(2, 2, 4)} & \multicolumn{1}{|l|}{18} & \multicolumn{1}{|l|}{9} &  \multicolumn{1}{|c|}{$0.5 \pm 0.1
$} & \multicolumn{1}{|c||}{$0.56 \pm
0.07$} 
\\
\hline
\multicolumn{1}{||c|}{(2, 3, 4)} & \multicolumn{1}{|l|}{25} & \multicolumn{1}{|l|}{10} &  \multicolumn{1}{|c|}{$0.4 \pm 0.2
$} & \multicolumn{1}{|c||}{$0.7 \pm
0.1$} 
\\
\hline
\multicolumn{1}{||c|}{(3, 0, 4)} & \multicolumn{1}{|l|}{16} & \multicolumn{1}{|l|}{8} &  \multicolumn{1}{|c|}{$0.0373 \pm 0.0005
$} & \multicolumn{1}{|c||}{$0.960 \pm
0.004$} 
\\
\hline
\multicolumn{1}{||c|}{(3, 1, 4)} & \multicolumn{1}{|l|}{12} & \multicolumn{1}{|l|}{9} &  \multicolumn{1}{|c|}{$0.9 \pm 0.1$} & \multicolumn{1}{|c||}{$0.49 \pm 0.05
$} 
\\
\hline
\multicolumn{1}{||c|}{(3, 2, 4)} & \multicolumn{1}{|l|}{20} & \multicolumn{1}{|l|}{10} &  \multicolumn{1}{|c|}{$0.48 \pm 0.09
$} & \multicolumn{1}{|c||}{$0.63 \pm
0.05$} 
\\
\hline
\multicolumn{1}{||c|}{(3, 3, 4)} & \multicolumn{1}{|l|}{28} & \multicolumn{1}{|l|}{11} &  \multicolumn{1}{|c|}{$0.7 \pm 0.1$} & \multicolumn{1}{|c||}{$0.5 \pm 0.1$} 
\\
\hline
\multicolumn{1}{||c|}{(4, 0, 4)} & \multicolumn{1}{|l|}{20} & \multicolumn{1}{|l|}{9} &  \multicolumn{1}{|c|}{$0.0389 \pm 0.0002
$} & \multicolumn{1}{|c||}{$0.959 \pm 0.003$} 
\\
\hline
\end{tabular} 
}
\caption{\textbf{Average Precision and KL values for various QNBM network structures.} Here, we provide the average results with their associated standard deviations across 5 training seeded runs. For a given 3-layer QNBM network with the structure $(N_{in}, N_{hid}, N_{out})$, we provide the number of tunable parameters in the circuit, the number of qubits required, the output precision for $10^4$ samples, and the KL divergence from the computed model output probability distribution. Ideal model performance is reached when we achieve  $P = 1$ and $KL = 0$. For various $N_{out}$, we see that optimal structures occur when $N_{hid} = 0$ and $N_{in} = N_{out}, N_{out} - 1$, indicating that hidden layers may not be advantageous to the performance of the model at small-scales.} 
\label{table:network_span}
\end{table}

\subsection{Investigating Non-Linearity }\label{sec3:non-linear}

\begin{figure*}[bht]
\subfloat[\label{sfig:model_comparison-a}]{%
  \includegraphics[width=0.25\linewidth]{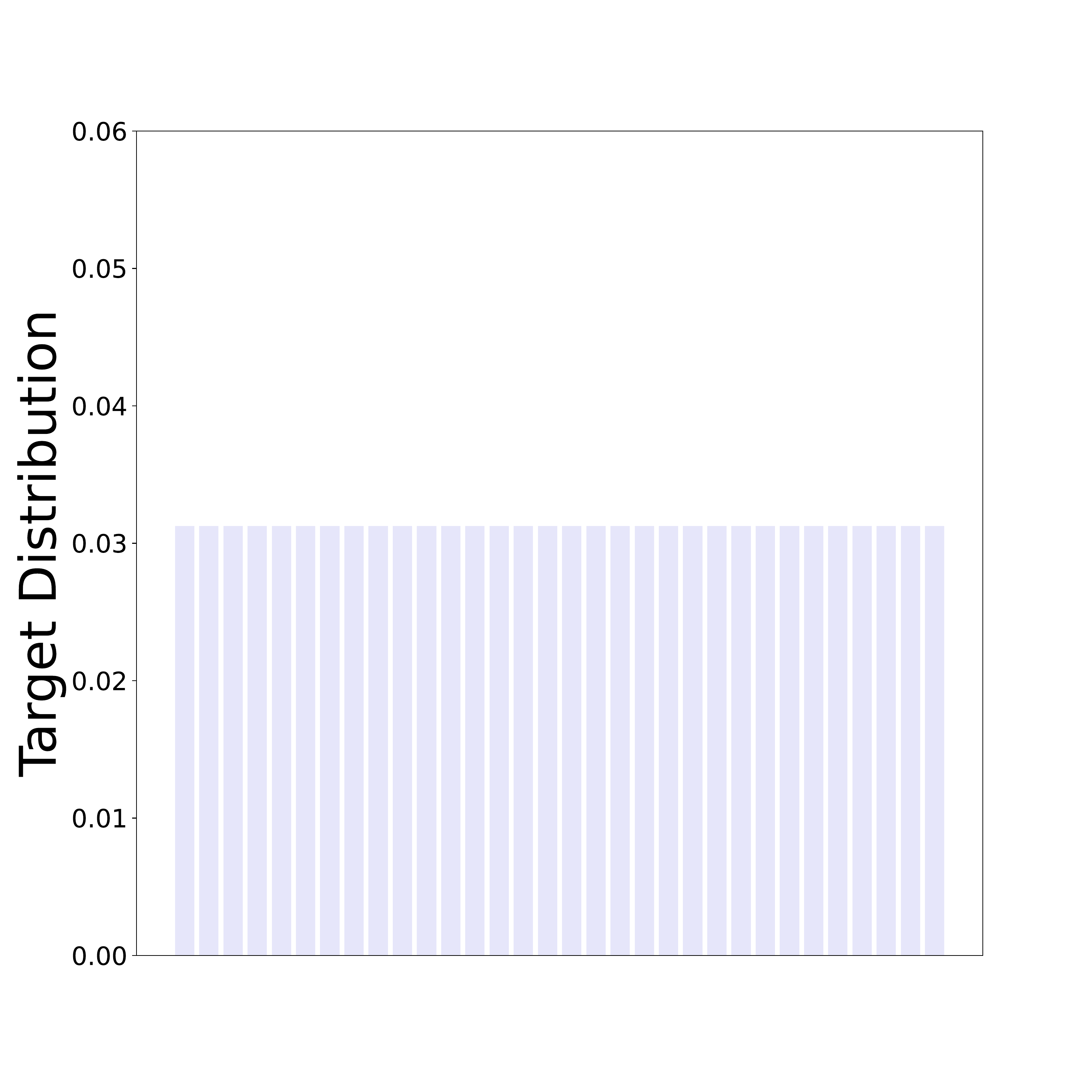}%
}
\subfloat[\label{sfig:model_comparison-b}]{%
  \includegraphics[width=0.25\linewidth]{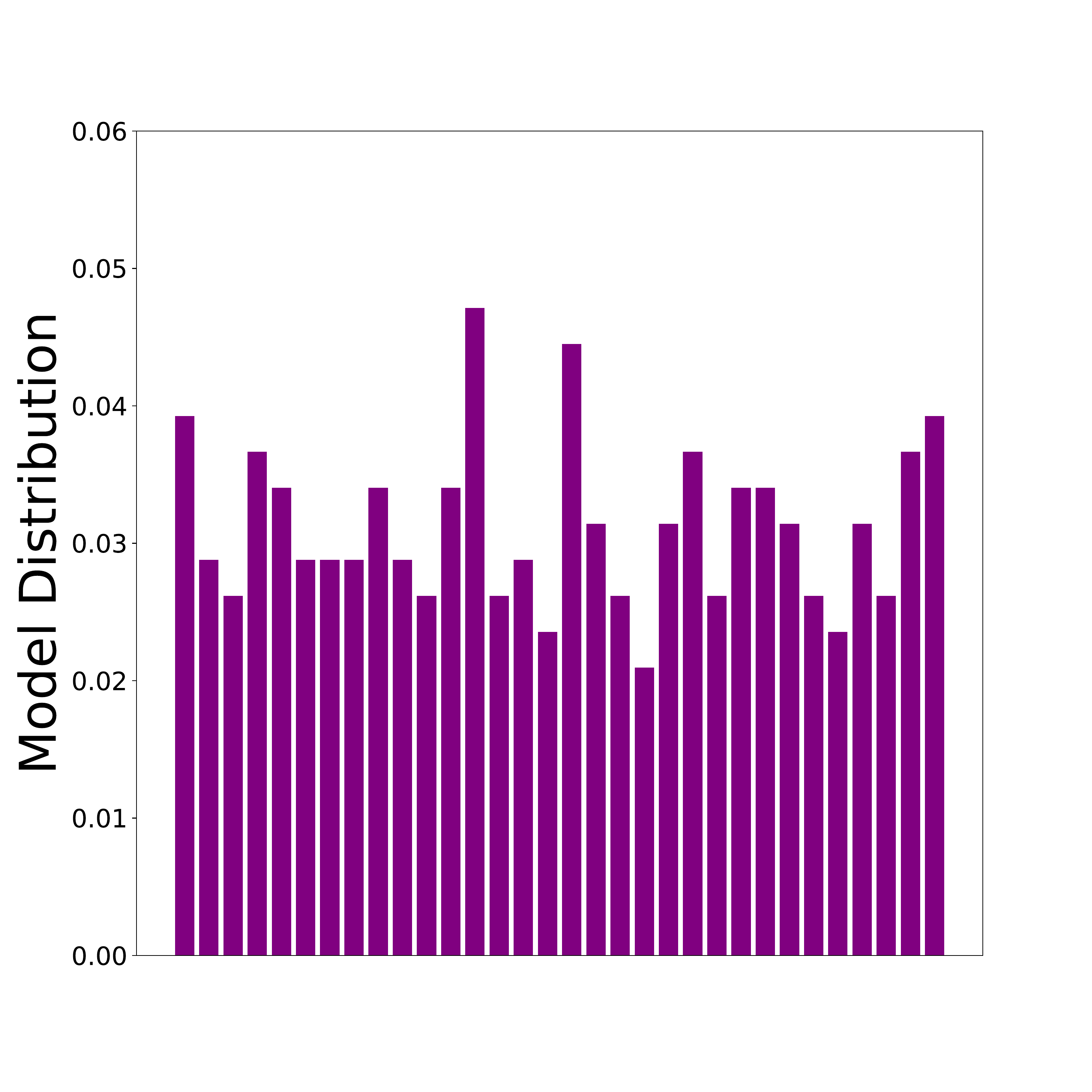}%
}
\subfloat[\label{sfig:model_comparison-c}]{%
  \includegraphics[width=0.25\linewidth]{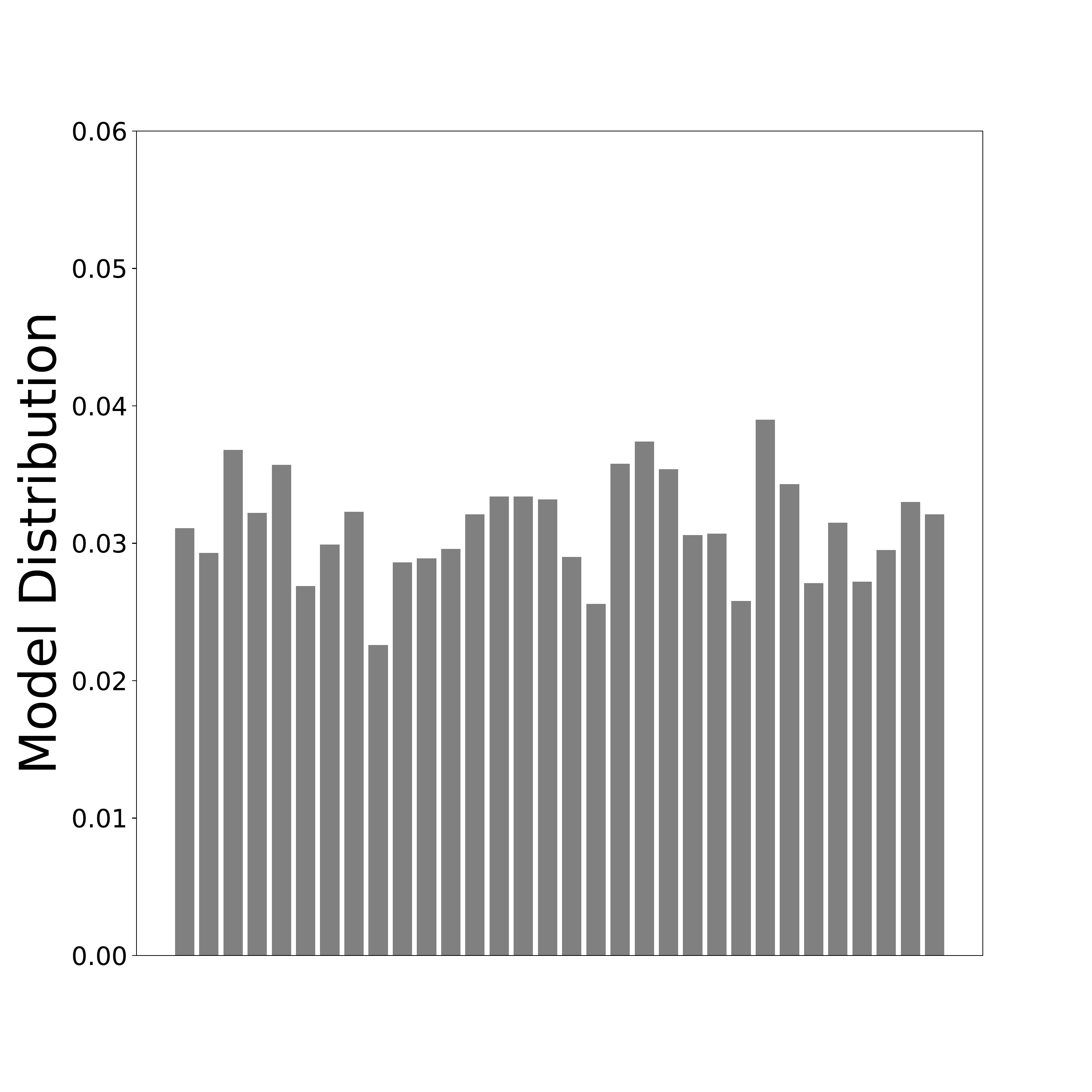}%
}
\subfloat[\label{sfig:model_comparison-d}]{%
  \includegraphics[width=0.25\linewidth]{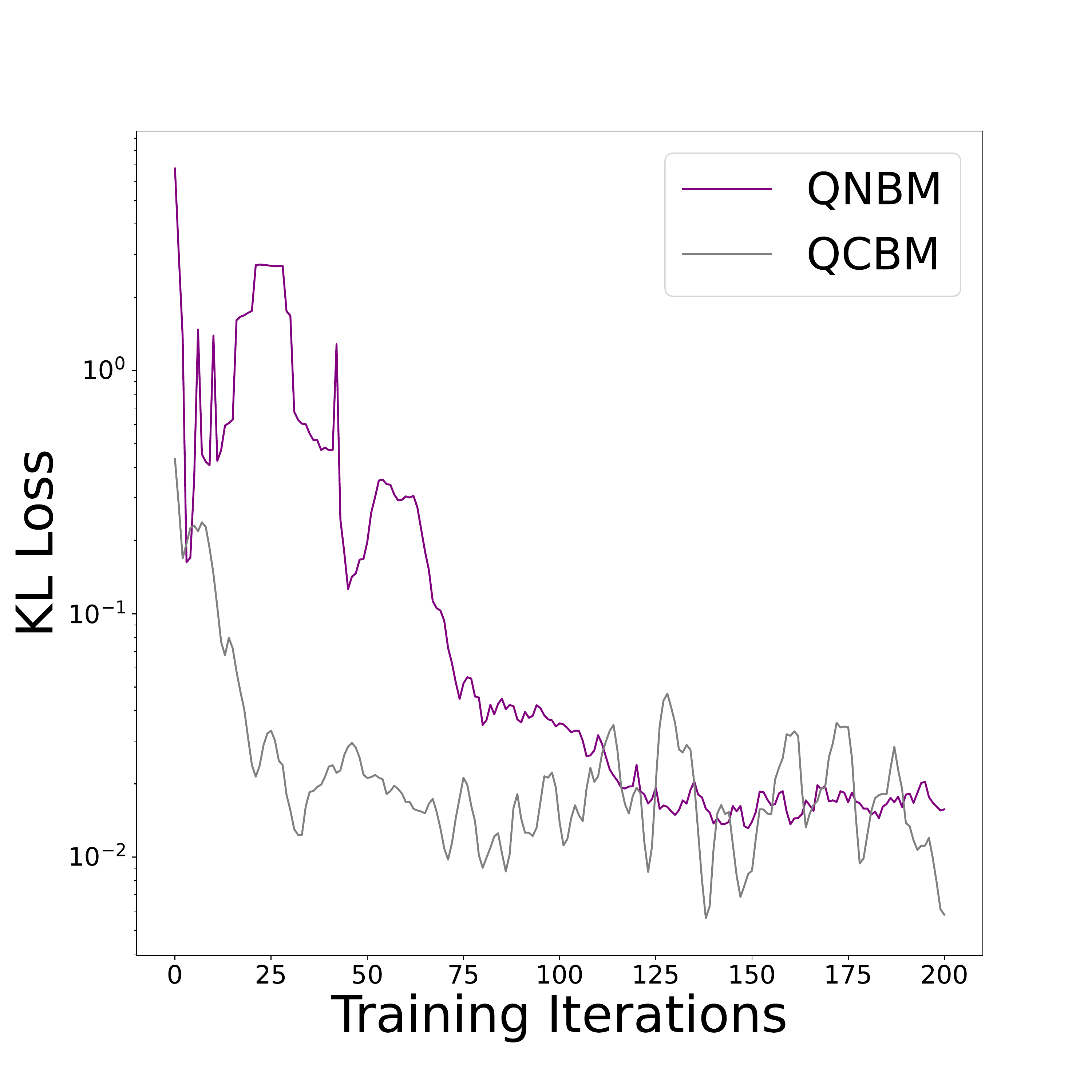}%
} \\
\subfloat[\label{sfig:model_comparison-e}]{%
  \includegraphics[width=0.25\linewidth]{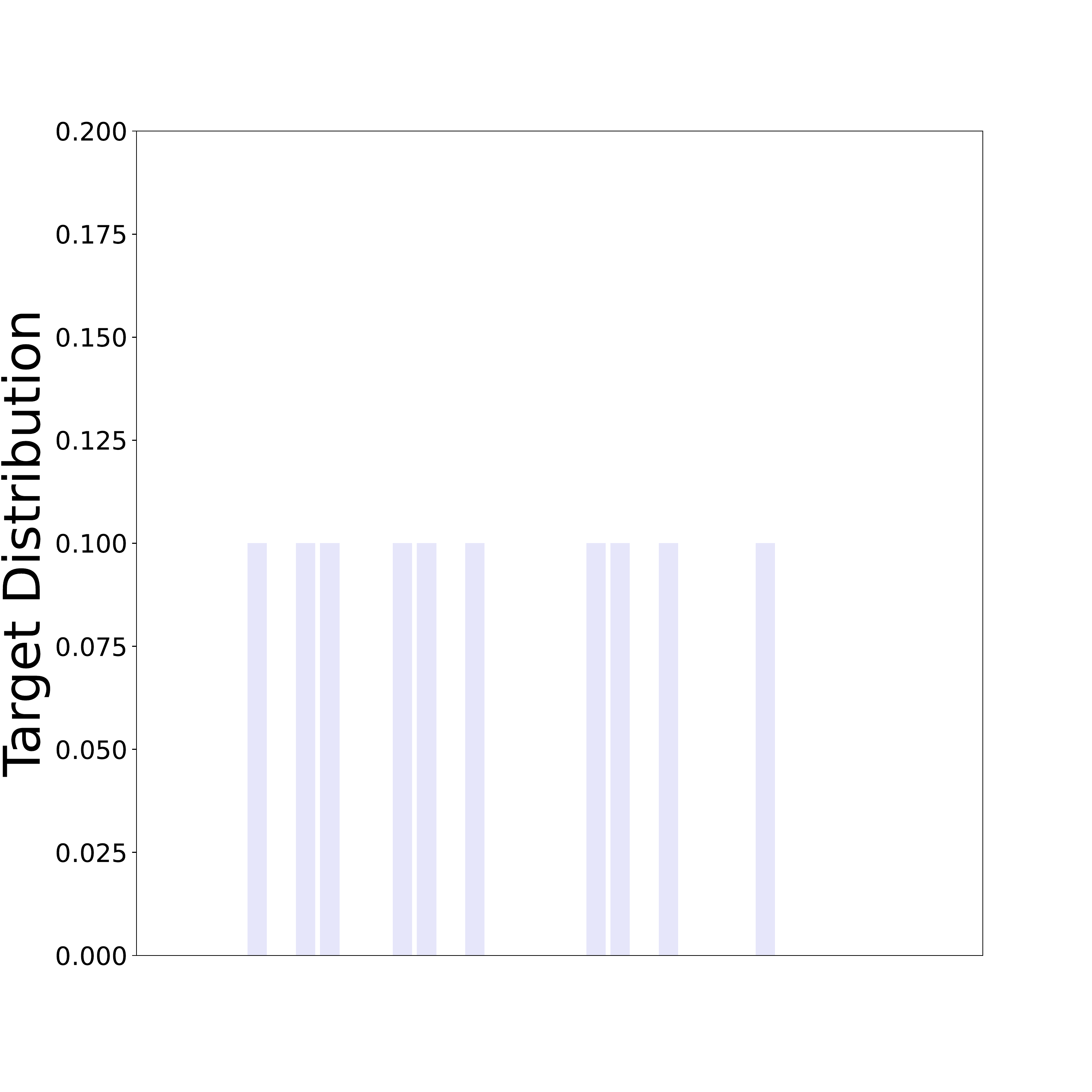}%
}
\subfloat[ \label{sfig:model_comparison-f}]{%
  \includegraphics[width=0.25\linewidth]{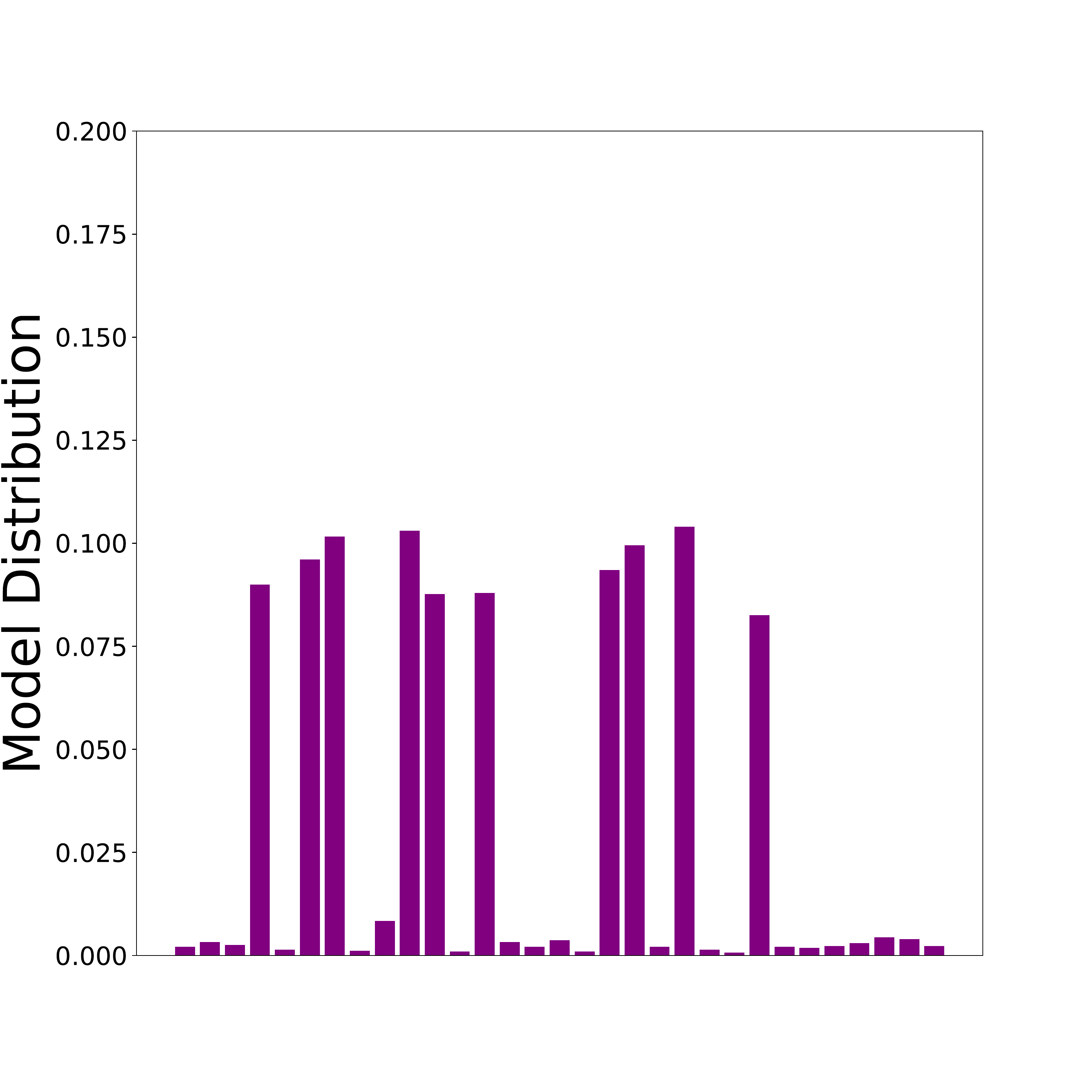}%
}
\subfloat[ \label{sfig:model_comparison-g}]{%
  \includegraphics[width=0.25\linewidth]{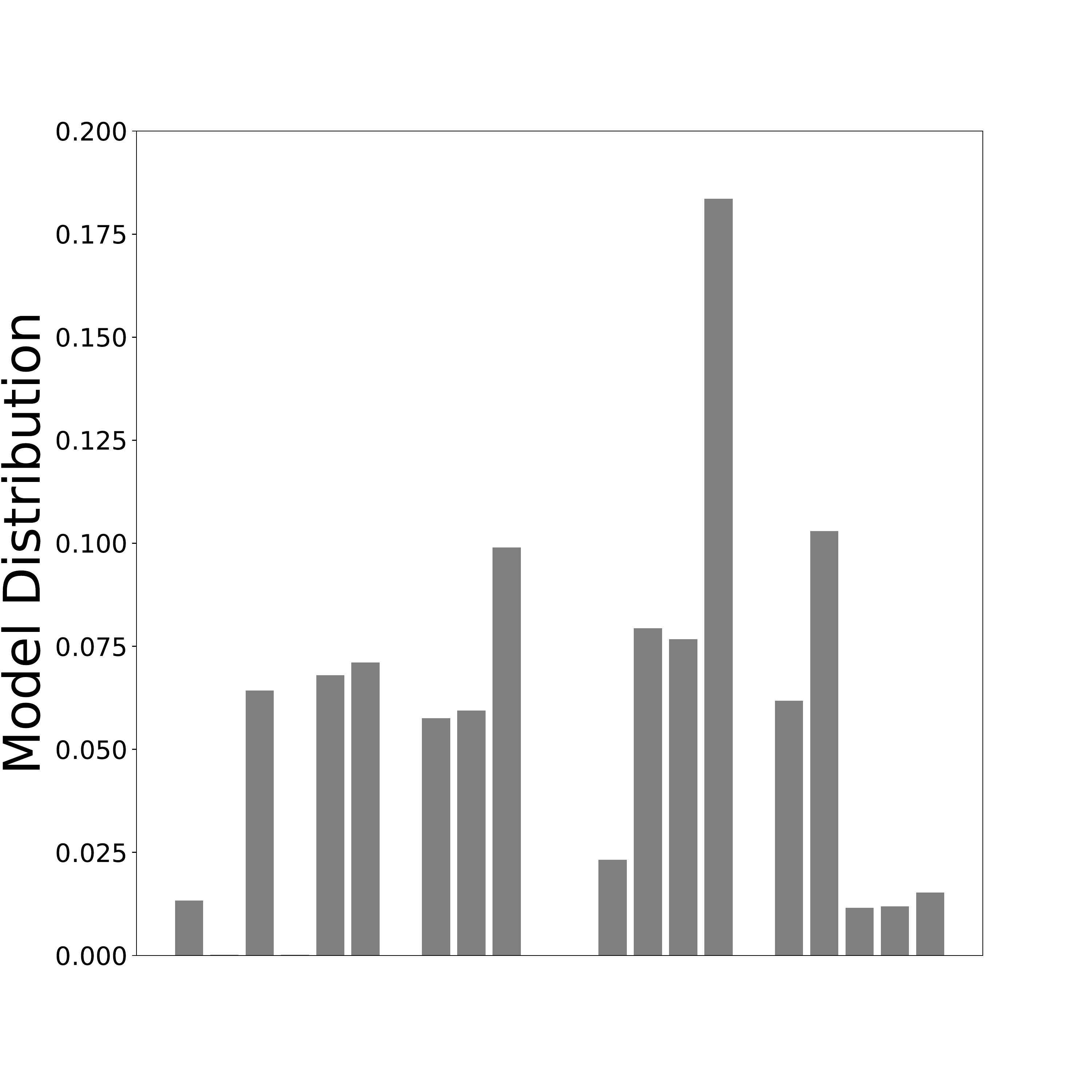}%
}
\subfloat[\label{sfig:model_comparison-h}]{%
  \includegraphics[width=0.25\linewidth]{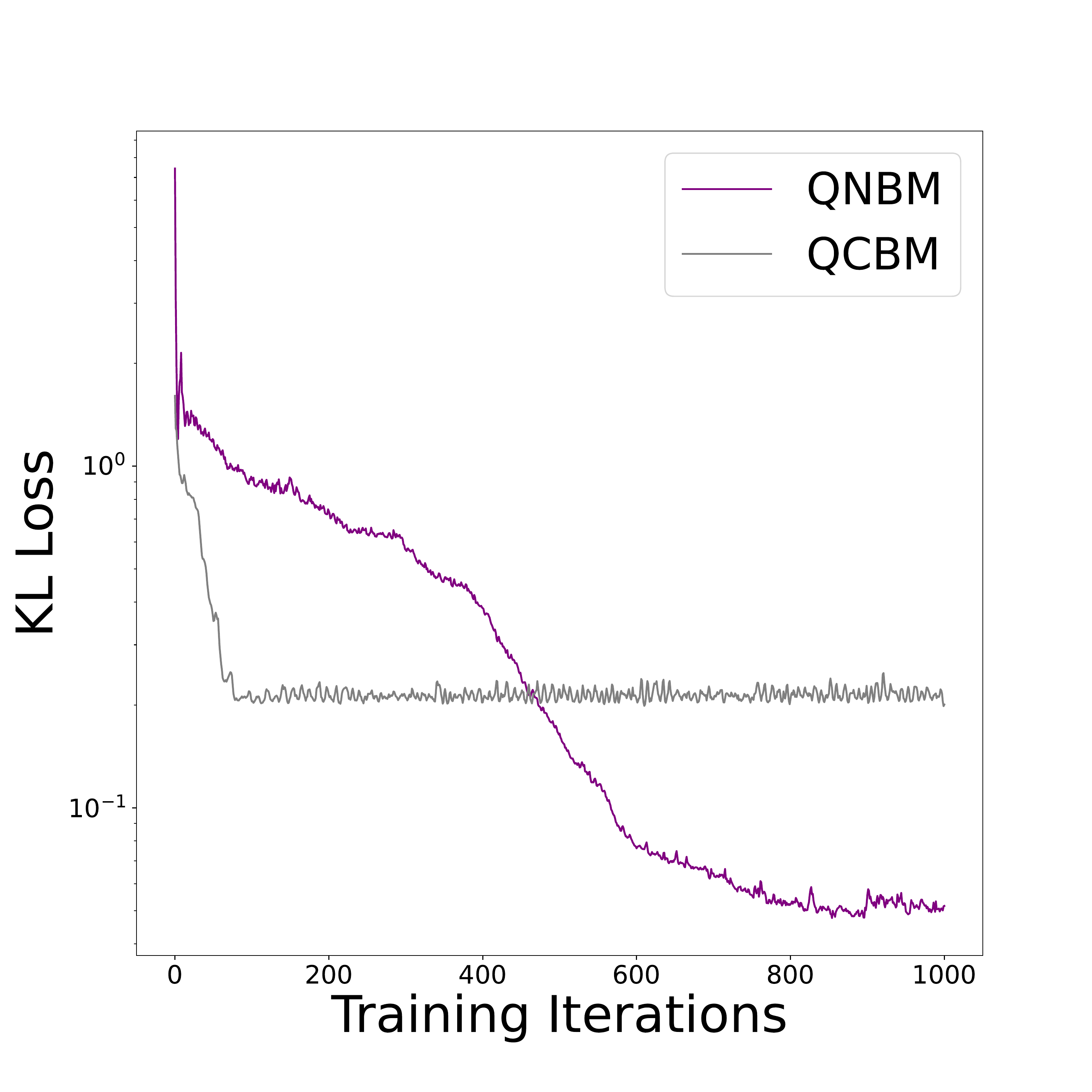}%
}
\caption{\textbf{A visual comparison between the Non-Linear QNBM and Linear QCBM in modeling uniform and constrained target distributions.} All probability histograms contain ordered binary bitstring samples on the x-axis. (a) The target uniform distribution. (b) The QNBM model output with $10^4$ samples. (c) The QCBM model output with $10^4$ samples. (d) The KL divergence loss throughout training for the QNBM and the QCBM. (e) The cardinality constrained target distribution. (f) The QNBM model output with $10^4$ samples. (g) The QCBM model output with $10^4$ samples. (h) The KL Divergence loss throughout training for the QNBM and the QCBM. We see that the QCBM and the QNBM are both able to learn the trivial uniform distribution well, where the QCBM even outperforms the QNBM in achieving the lowest KL value in the fewest iteration steps. However, we see that with added complexity, the QNBM is able to outperform the QCBM and continue learning after the QCBM reaches its plateau in training, resulting in almost 3x smaller error rate.}
\label{sfig:model_comparison}
\end{figure*}

In addition to investigating the network design, we study how introducing non-linearity into the quantum circuit model enhances the model's ability to learn a more complex, cardinality-constrained distribution. To achieve this, we compare our non-linear QNBM to a linear QCBM with the same $N_{out} = 5$, and a comparable number of tunable parameters: $P_{\text{num}} = 25$ for the QNBM with the topology $(4, 0, 5)$ and $P_{\text{num}} = 20$ for a QCBM with an all to all connected 1 single and 1 entangling layer. We select this QCBM topology as it has demonstrated great success in the literature \cite{Benedetti_2019}. Both models are trained with an Adam gradient-based optimizer with a learning rate $\alpha = 0.2$ and a step size $\epsilon = 0.1$ for the same number of iterations. We aim to keep the number of quantum and classical resources similar across both models to evaluate the models on equal ground to the best of our ability. We conduct 5 runs of each model with various training seeds, and use the best training for each model in our results, where we take the best training to be the model whose last loss is the lowest across the seed trials. All simulations are run on IBMQ's Qiskit Aer Simulator with $N_{\text{shots}} = 10^4$ shots. 

We evaluate both models for their ability to produce two different probability distributions: one trivial distribution and one with added complexity. We take the trivial distribution to be one that is completely uniform over the search space $\mathcal{U}$, such that: 
\begin{equation}\label{search_prob}
P_{\mathcal{U}}(x) = \frac {1} {|\mathcal{U}|}, \forall x \in \mathcal{U}. 
\end{equation}

We take the second distribution to be the same as \eqref{data_prob} with $N_{out} = 5$ and $c = 2$, as introducing a cardinality constraint adds more complexity to the distribution by enhancing the difficulty of the learning problem. In \figref{sfig:model_comparison}, we show the performance comparison of the QNBM to the QCBM on the two distributions by displaying their model output probability distributions produced at the end of training, and their KL training curves. We see that for the uniform target, both the QCBM and the QNBM are able to accurately model the probability distribution (QCBM: $KL = 0.0058$ QNBM: $ KL = 0.0157$). Additionally, we see that throughout training, the QCBM takes fewer iterations to converge. However, when adding a constraint to the distribution and thus increasing the complexity, we see that the non-linear QNBM outperforms the linear QCBM with almost 3 times smaller error rate $1-P$ (QNBM: 0.05, QCBM: 0.14). In monitoring the loss throughout training, we see that the QCBM hits a plateau earlier on in training, while the QNBM is able to continue on in the training process and generate higher quality samples. 

In addition to our comparison where we keep the allocated resources as similar as possible, we note that adding a second layer of single-qubit and two-qubit parameterized gates to the QCBM increases its performance as shown in the Appendix (\secref{sfig:qcbm_2layers}). However, this requires twice as many tunable parameters (40, more than the total number of possible bitstrings), and still achieves a two times larger error rate than the QNBM.

To further support that this training enhancement is due to non-linearity rather than just the neural network structure of the quantum circuit, we additionally train a "linearized" QNBM where Quantum Neuron subroutines are replaced with linear Pauli-Y rotations. In \figref{sfig:qnbm_linear}, we show that this model also struggles to learn the cardinality constrained distribution as well as the QNBM, suggesting that non-linear activation functions are essential to the good performance of the QNBM.

\section{Discussion and Outlook}\label{s:outlook}
In this work, we have shown that quantum generative models can benefit from quantum state evolution which is non-linear. We have introduced and put forth an initial investigation of a model that inherently contains this non-linearity through a neural network structure. We have shown that in the small-scale regime, it outperforms the QCBM - one of the most prominent quantum generative models - in learning a challenging probability distribution. We aimed to make the quantum circuit comparison as fair and unambiguous as possible - with similar circuit depth and number of parameters. Additionally, we kept the classical resources fixed for each training.

The only significant resource discrepancy in the model comparison is in the qubit number. While QCBMs need as many qubits as there are bits in the target probability distribution, QNBMs have a qubit overhead which grows with the number of non-output neurons in the network. If its good performance on complex datasets is transferable to larger-scale problems, the QNBM could fit within the Born Machine family as a model that exchanges “memory” resources – circuit width – for “time” resources – number of optimizer iterations to learn a given probability distribution. Our numerics suggest that this trade-off is, at least for small-scale examples, efficient: the best networks seem to have no hidden layer and no more input neurons than output neurons, hence the overhead should grow at most linearly with the number of output bits.

For future work, we see two main avenues of research. The first relates to a more detailed understanding of the QNBM’s capabilities. Primarily, it would be useful to investigate whether the QNBM performs well on other types of probability distributions and can generalize to unseen data, rather than just memorize distributions \cite{kgili_gen, gen_mmd, gili_generalize}. It would also be useful to investigate how much the high two-qubit gate count of the QNBM hinders its performance on near-term hardware. In addition, one could also explore whether any useful information is stored in the non-output qubits post-sampling, similarly to how hidden layers of classical neural networks sometimes self-organize in an interpretable fashion to learn different aspects of data. Lastly, we leave for future work the exploration of analytical cost gradients for QNBMs. As the model contains mid-circuit measurements, one cannot naively apply the parameter shift rule \cite{Mitarai_2018} when calculating gradients of observable expectation values with respect to the weights and biases, but other tricks may be possible.

The second avenue relates to a better and potentially analytical understanding of why QNBMs have any performance advantages. Our work leads us to believe that it is the combination of non-linearity and the inherent structure that yields these advantages, rather than just one of the two. It would be interesting to see whether its performance could be explained by its limited expressivity. The QNBM has a very strong inductive bias in state space due to effectively only applying Pauli-Y rotations, resulting in a model that for uniformly random parameters creates states that are very much not Haar-random. We know high expressivity can lead to barren plateau phenomena, where gradients of the cost exponentially vanish \cite{McClean2018bp}. Therefore, we think the restricted expressivity of the QNBM deserves further investigation as a potential work-around against barren plateaus.

As quantum computing technology improves and larger-scale applications become possible, performance patterns of various quantum generative models could drastically differ from those observed in small-scale demonstrations. Based on our work, we do not place the QNBM as the definitive candidate to achieve quantum advantage, but rather as one of many formidable candidates with its own unique advantages. Much more importantly, we hope this work encourages the quantum machine learning community to consider non-linearity as another resource worth utilizing when designing generative models.

\begin{acknowledgments} 
The authors would like to recognize the Army Research Office (ARO) for providing funding through a QuaCGR PhD Fellowship. This work was supported by the U.S. Army Research Office (contract W911NF-20-1-0038) and UKRI (MR/S03238X/1).  Additionally, the authors would like to recognize William Simon for feedback on an early version of the manuscript. Lastly, the authors would like to acknowledge the places that inspired this work: Oxford, UK; Muscat, Oman; Lisbon, Portugal; Berlin, Germany; Lake Como, Italy; and Chicago, Boston, Greensboro, USA. Thank you to all of the local people who became a part of the journey. 
\end{acknowledgments}

\bibliography{quantum.bib}
\onecolumngrid
\pagebreak 
\section{Appendix}\label{s:appendix}

\subsection{Supplemental Support}\label{appendix:support}

\begin{figure*}[bht]
\subfloat[\label{sfig:uniform_dist}]{%
  \includegraphics[scale = 0.2]{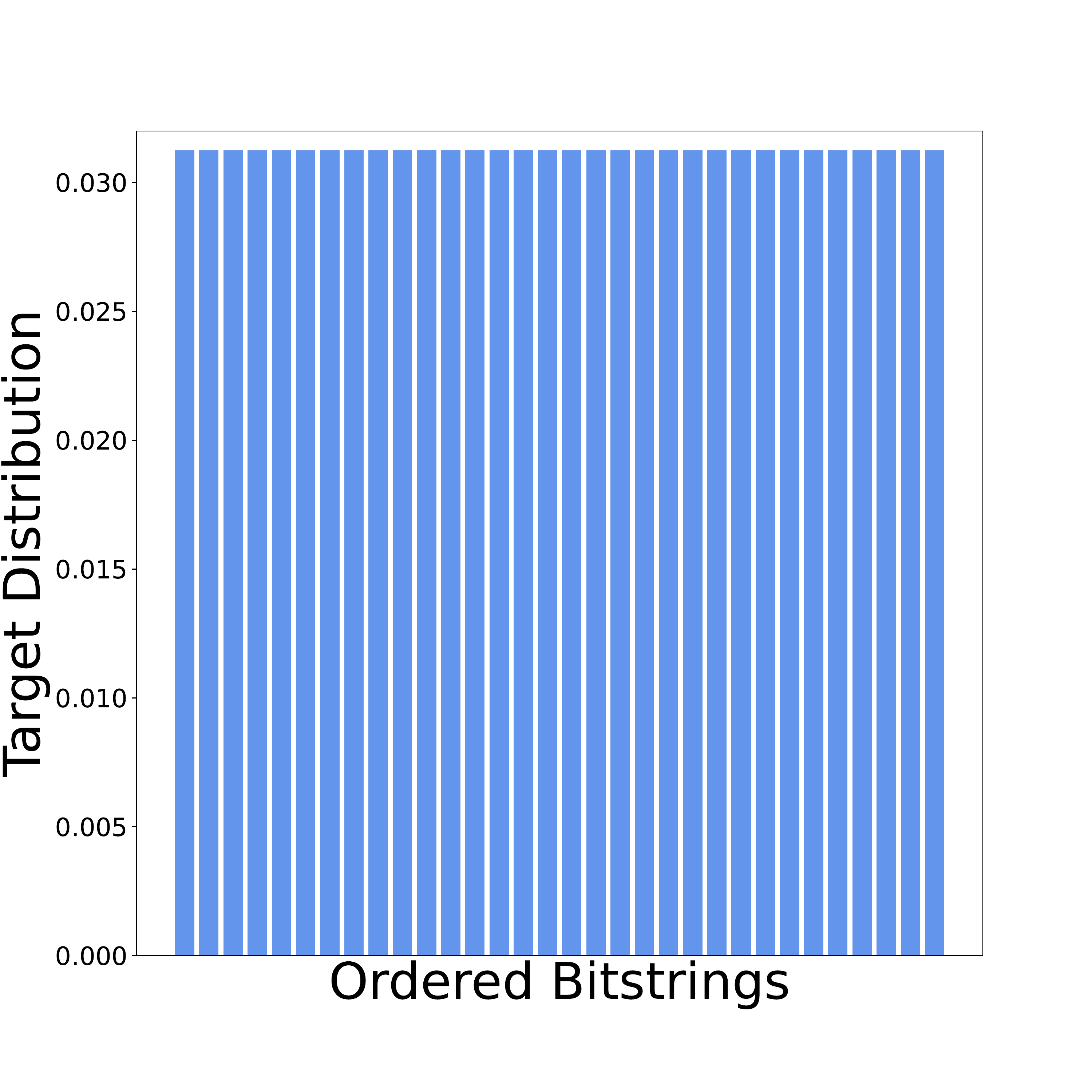}%
}
\subfloat[\label{sfig:qcbm_2layer-a}]{%
  \includegraphics[scale = 0.2]{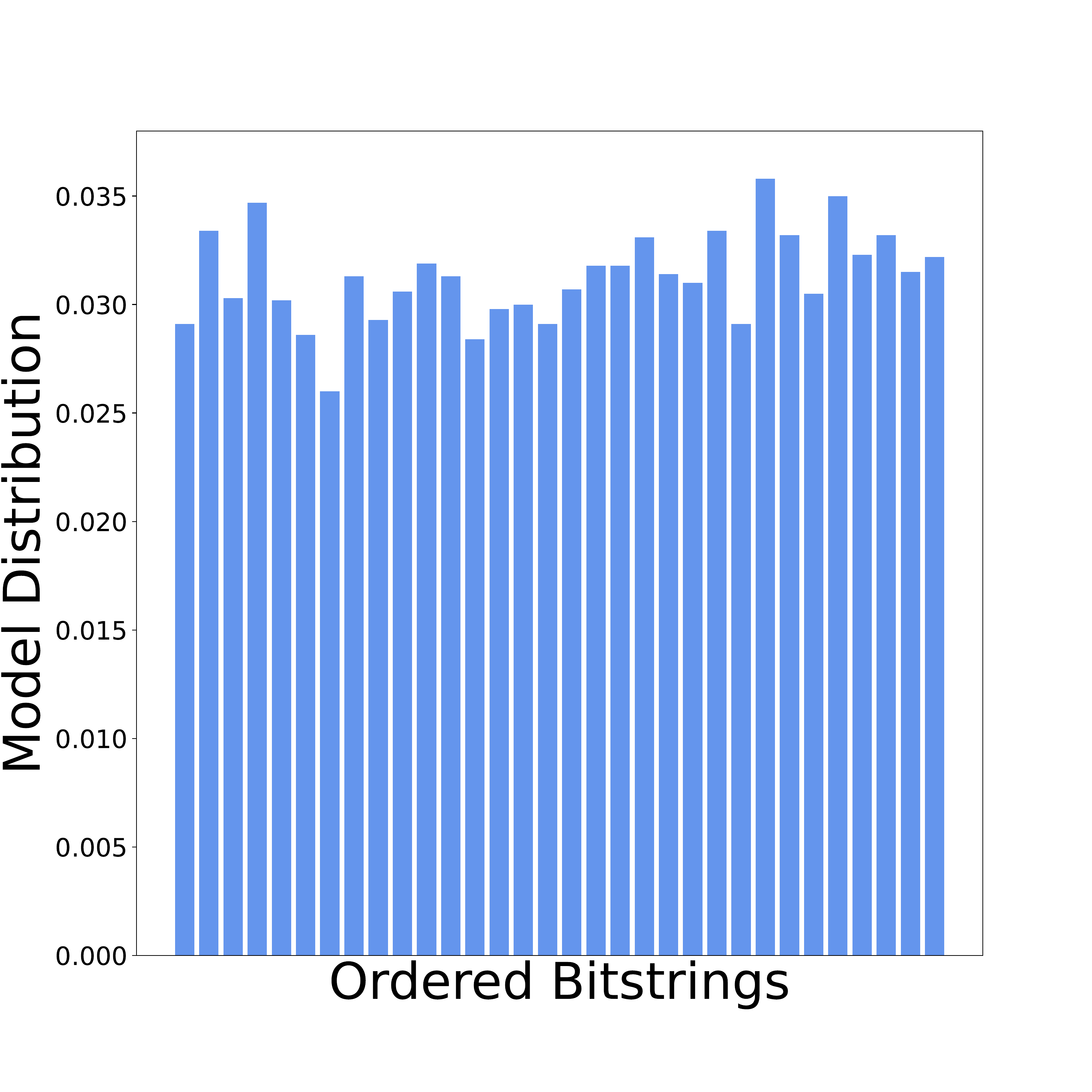}%
} \\
\subfloat[\label{sfig:constrained_dist}]{%
  \includegraphics[scale = 0.2]{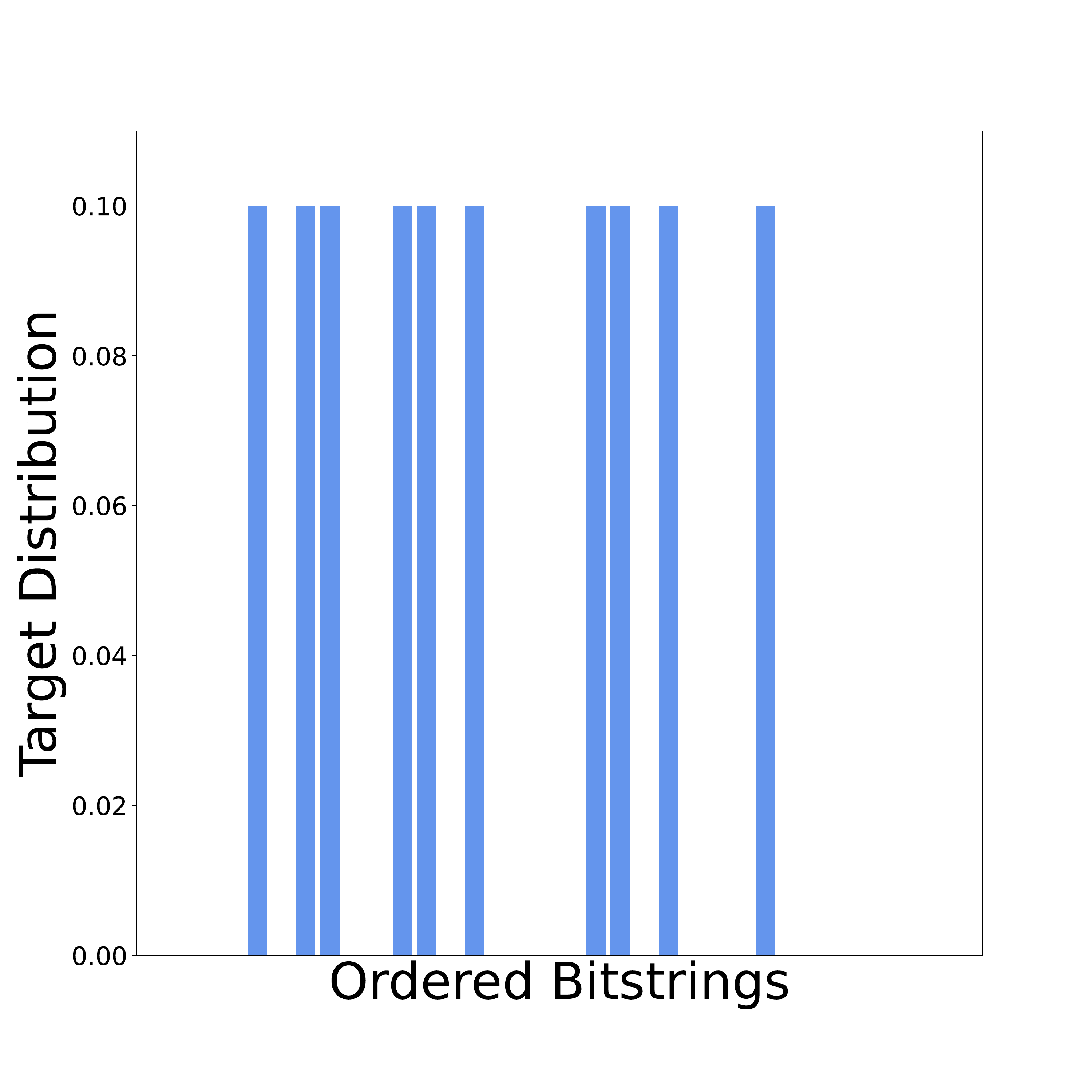}%
}
\subfloat[\label{sfig:qcbm_2layer-b}]{%
  \includegraphics[scale = 0.2]{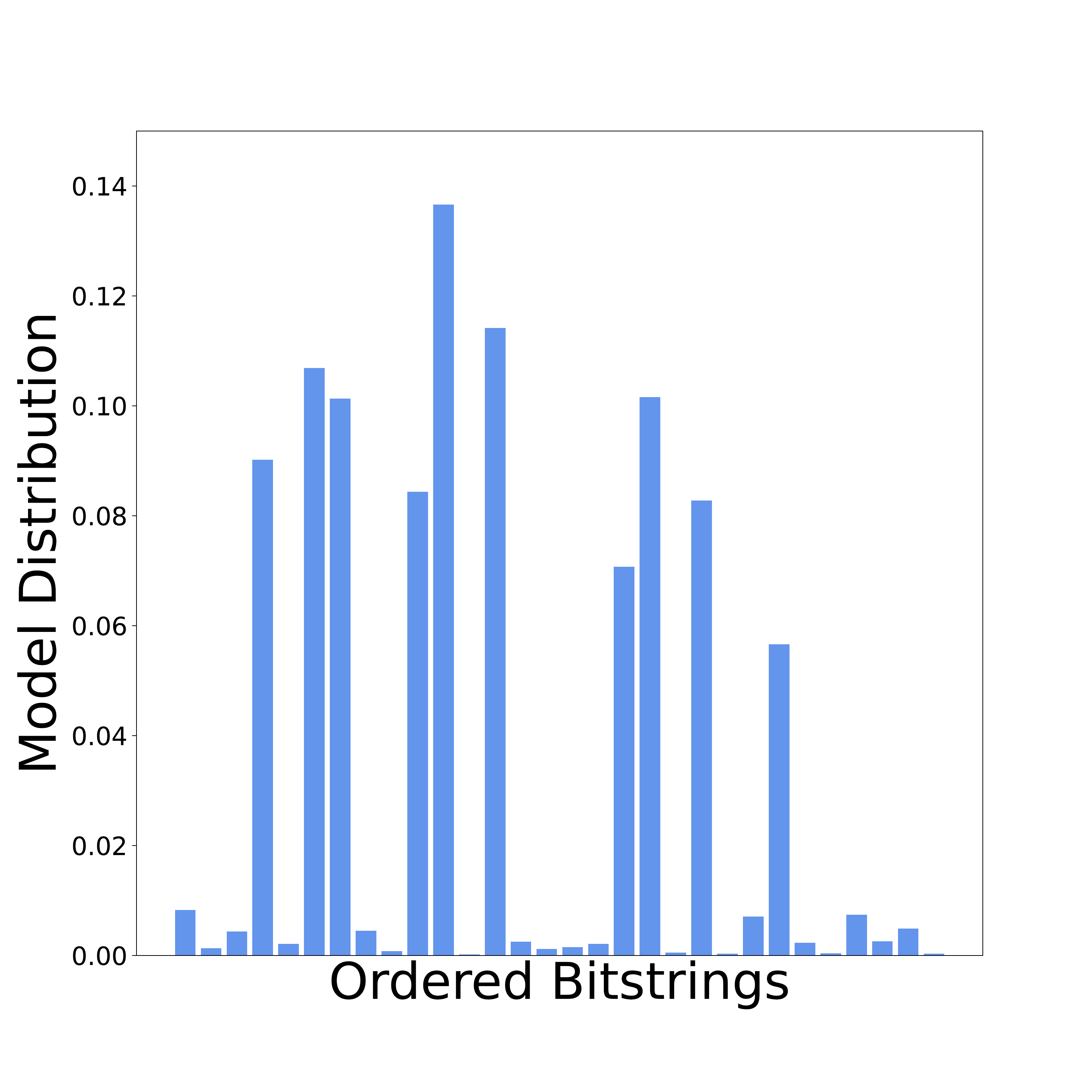}%
}
\caption{\textbf{The QCBM model performance when given more expressivity.} (a) The uniform target probability distribution. (b) The 2-layer trained QCBM output probability distribution with $Q = 10k$ samples. (c) The cardinality constrained $c = 2$ target probability distribution. (d) The 2-layer trained QCBM output probability distribution with $Q = 10k$ samples. We see that for the uniform distribution, the QCBM is able to reproduce the target distribution well, but for the constrained distribution, the QCBM performs very poorly. Overall, the expressibility does not provide an enhancement for the QCBM to potentially outperform the QNBM.}
\label{sfig:qcbm_2layers}
\end{figure*}

Here, we display the probability histograms for the $N_{out} = 5$ QCBM with multiple layers, specifically 2 single qubit gate layers and 2 entangling layers of the topology mentioned in \secref{sec2: qcbm}. In \figref{sfig:qcbm_2layers}, we compare these histograms with the target probability distributions for the uniformly distributed case and the cardinality constrained $c = 2$ case in order to detect if adding more expressibility to the circuit increases the training capabilities. As always, we use the Adam gradient-based optimizer with learning rate $\alpha = 0.2$ and step size $\epsilon = 0.1$, and we run the optimizer for 1000 iterations for the cardinality constrained distribution and for 200 iterations for the uniform distribution. While for the uniform distribution, we see that the model is able to learn the target distribution either just as well or better than the QCBM with 1 layer (single plus entangling), the cardinality constrained distribution becomes more difficult to learn with additional layers. In the case of the best training (the lowest KL across 5 seed training runs), we see that the model does not train well ($KL = 1.08$) and precision performance decreases ($P = 0.35$). Hence, increasing the expressibility does not help the QCBM outperform the QNBM for this task. 

\begin{figure*}[bht]
\subfloat[\label{sfig:card_con_dist}]{%
  \includegraphics[scale = 0.2]{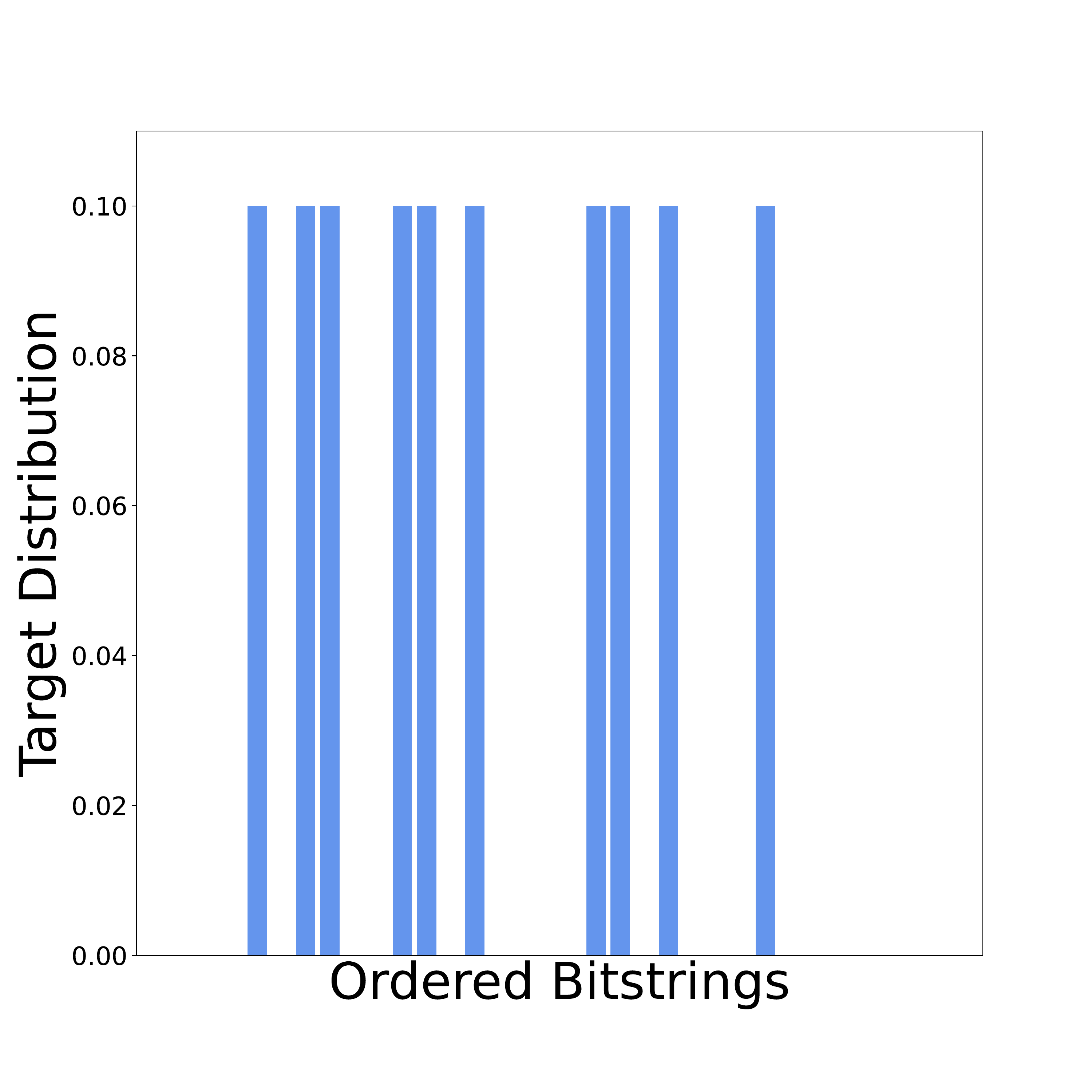}%
}
\subfloat[\label{sfig:qnbm-linear-a}]{%
  \includegraphics[scale = 0.2]{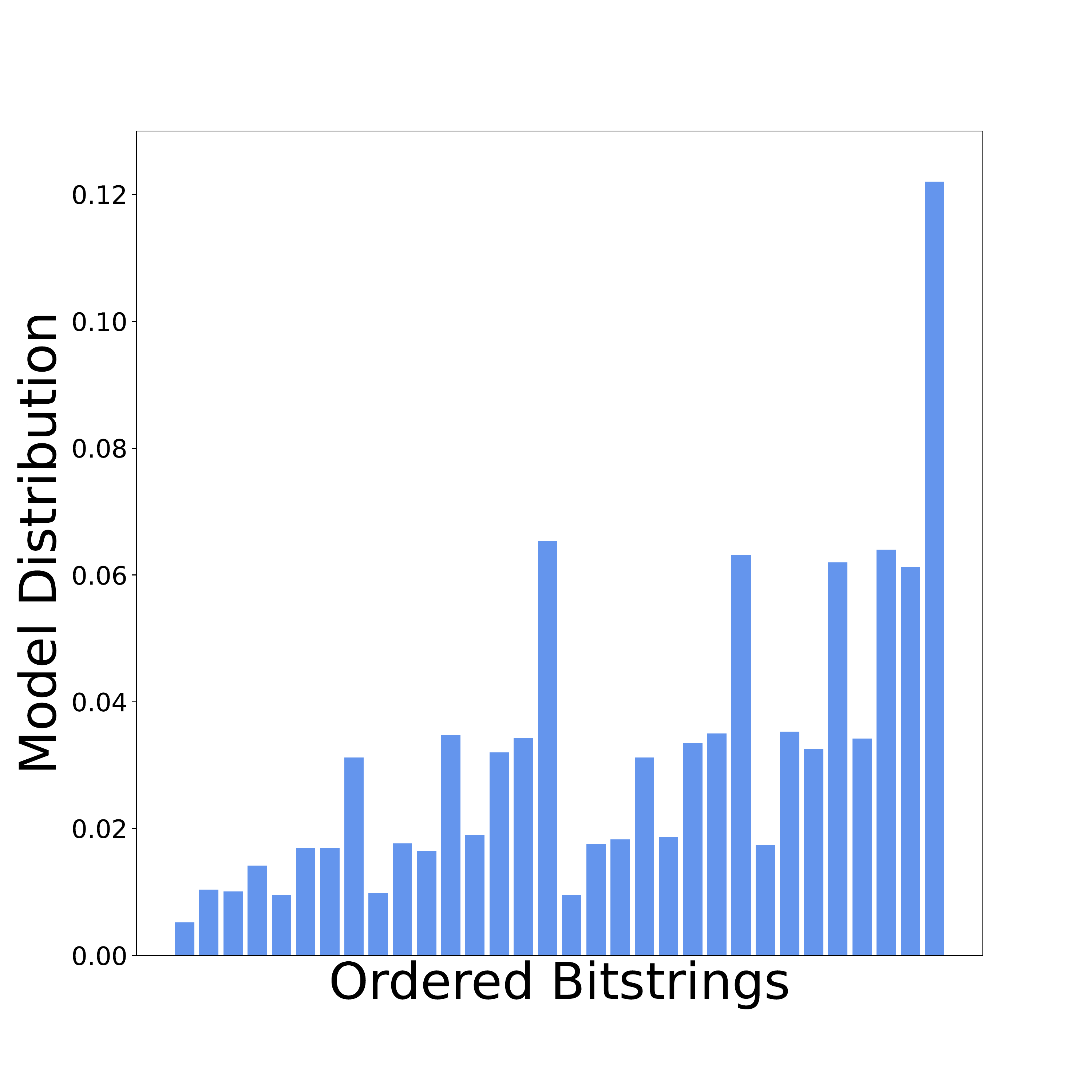}%
} 
\caption{\textbf{The linear QNBM model performance on the cardinality-constrained distribution.} (a) The cardinality constrained $c = 2$ target probability distribution. (b) The linear QNBM output probability distribution with $Q = 10k$ samples. We see that without the non-linearity, the model greatly suffers in training and learning the target distribution. Therefore, we consider the non-linearity to be a necessary resource in addition to the ansatz structure.}
\label{sfig:qnbm_linear}
\end{figure*}

Additionally, we train a "linearized" version of the QNBM, which has the same structure as in \figref{fig:qnbm}, but with every RUS Quantum Neuron subroutine replaced by a controlled Pauli-Y rotation. The only effect of this is to substitute non-linear rotations $R_Y(2q(\theta))$ with linear rotations $R_Y(2\theta)$. We use the same cardinality-constrained distribution with $N_{out} = 5$ and $c = 2$ for the target. Again, we use the same optimizer, hyperparameters, and number of iterations as before. The final result is demonstrated in \figref{sfig:qnbm_linear}.

With this implementation, we are able to further support that it is not only the neuron structure that gives rise to the QNBM's good performance, but the combination of structure and the non-linearity. The model does not train well at all, and has poor precision when sampled post-training. Over 5 seed runs, the best model achieves a $KL = 2.15$ with a corresponding precision of $P = 0.173$. We thus conclude that for this neuron structure, non-linear transformations into the state evolution is a requirement to achieve good generative performance. 

\end{document}